\documentclass[runningheads]{llncs}
\usepackage[T1]{fontenc}
\usepackage{graphicx}
\usepackage{booktabs} 
\usepackage{url}
\usepackage{cite}
\usepackage{amsmath}
\usepackage{multirow}

\begin{document}
\title{Mapping Ecological Empathy: A Semantic Network Analysis of Player Perceptions in 3D Environmental Education Games\thanks{This research was funded by: GR024678 NSERC CREATE 2020 Immersive Technologies, Natural Sciences and Engineering Research Council of Canada; GR026895 SSHRC 2022 Okanagan WaterFutures: Experiential Games for Water Responsibility, Social Sciences and Humanities Research Council of Canada.}}

\titlerunning{ Mapping Ecological Empathy} 

\author{Yuanyuan Xu\inst{1} \and
Zhehao Sun\inst{1} \and
Chi Zhen\inst{1} \and
Yin-Shan Lin\inst{2} \and
Miles Thorogood\inst{1} \and
Megan Smith\inst{1} \and      
Patricia Lasserre\inst{3} \and
Aleksandra Dulic\inst{1}}      

\authorrunning{Y. Xu et al.} 

\institute{Faculty of Creative and Critical Studies, University of British Columbia,\\
Kelowna, BC V1V 1V7, Canada\\\and
Khoury College of Computer Sciences, Northeastern University,\\
Vancouver, BC V6B 1Z3, Canada \and
Department of Computer Science, Irving K. Barber Faculty of Science,\\
University of British Columbia, Okanagan, BC V1V 1V7, Canada}

\maketitle

\begin{abstract}
As the global climate crisis intensifies, 3D video games have emerged as powerful, interactive simulations for Environmental Education (EE). However, empirical assessment of their pedagogical efficacy remains epistemologically challenged. Traditional evaluation metrics, such as pre-post surveys, often suffer from response bias and fail to capture the nuanced, emergent psychological shifts players experience during gameplay. This paper proposes a novel, non-intrusive approach: utilizing Semantic Network Analysis (SNA) to map the "unsupervised" cognitive structures of players. We scraped and qualitatively filtered 1,825 rich-text user reviews from Steam for two distinct titles representing opposing ecological philosophies: \emph{Eco} (anthropocentric systemic management) and \emph{WolfQuest} (biocentric embodied survival). By constructing co-occurrence networks and calculating topological metrics, we visualized the divergence in how players conceptualize human-nature relationships. Results indicate a fundamental pedagogical split: \emph{Eco} promotes "Socio-Political Cognition," where environmental challenges are framed as legislative and economic frictions; conversely, \emph{WolfQuest} fosters "Effective Empathy," where players internalize the fragility of life through the vulnerability of the avatar. We argue that semantic topology offers a rigorous methodological tool for serious games assessment, revealing that effective environmental education requires a strategic tension between systemic logic and emotional resonance.

\end{abstract}

\section{Introduction}
Digital games are increasingly recognized not merely as entertainment, but as complex semiotic domains capable of actively shaping players' environmental literacy and attitudes \cite{gee2003, squire2011}. Unlike passive media (documentaries or textbooks), 3D environmental games position players as active agents within dynamic systems, forcing them to negotiate the consequences of resource extraction, survival strategies, and ecological balance \cite{wu2019}. The defining characteristic of these "Green Games" is their potential to transform abstract, high-level climate data into tangible, experiential learning moments \cite{flood2020}.

However, a critical methodological gap exists in the evaluation of these interventions \cite{liarakou2023, sherry2022}. How do researchers strictly differentiate between a player enjoying the \emph{mechanics} of a simulation and a player genuinely comprehending the \emph{underlying ecological message}? Traditional evaluation methods predominantly rely on explicit feedback mechanisms, such as Likert-scale surveys or interviews \cite{kao2017, zhang2024}. These methods are intrusive, disrupting the "Magic Circle" of gameplay \cite{huizinga1955, salen2004}, and often measure short-term declarative knowledge rather than long-term attitudinal shifts \cite{plass2015}. Recent advances in game-based learning assessment suggest that implicit, stealth assessment approaches, embedded within gameplay mechanics, may offer a more ecologically valid alternative , capturing behavioral markers that reflect genuine engagement with environmental concepts without interrupting the ludic experience \cite{ difrancesco2021}.

This paper aims to answer two pivotal research questions regarding the design of environmental pedagogy:
\begin{enumerate}
    \item \textbf{Structural Divergence:} How does the semantic topology of player discourse differ between games focusing on \emph{Macro-Management} (Eco) and those focusing on \emph{Micro-Survival} (WolfQuest)?
    \item \textbf{Bridging Concepts:} What specific lexical nodes bridge the gap between "Gameplay Fun" and "Ecological Education" in the player's mind?
\end{enumerate}

\section{Theoretical Framework} \label{sec:theory}
This research integrates dual theories from Environmental Psychology and Computational Social Science to construct a robust model for analyzing game-based learning \cite{arnab2015}.

\subsection{The Taxonomy of Agency: Anthropocentric vs. Biocentric Agencies}
Environmental Education (EE) is often paralyzed by a dichotomy between cognitive understanding and emotional connection \cite{russ2015}. We utilize this tension to categorize our case studies into two opposing "Epistemic Models" :

\begin{itemize}
    \item \textbf{System Thinking (The Anthropocentric Model):} This approach emphasizes the understanding of non-linear interdependencies, feedback loops, and resource flows \cite{sterling2010, sterman2018}. Games like \emph{Eco} embody this model by simulating the "Tragedy of the Commons" \cite{hardin1968, ostrom2015}. Here, the environment is framed as a distinct object to be managed, regulated, and optimized by human intellect \cite{kopnina2014}. The primary challenge is social coordination against scarcity \cite{kollmuss2002}.
    
    \item \textbf{Embodied Empathy (The Biocentric Model):} Conversely, theories of Embodied Cognition suggest that learning occurs through sensorimotor coupling with the environment \cite{barsalou2008, wilson2002}. By controlling an animal avatar in games like \emph{WolfQuest}, players undergo a "Perspective Taking" exercise tailored to the "More-Than-Human" world \cite{kahn2009, taylor2017}. In this model, the environment is not a resource to be managed, but a habitat to be endured. The primary pedagogy is the cultivation of an "Ethics of Care" through vulnerability \cite{noddings2013, tronto1993}.
\end{itemize}

Semantic Network Analysis allows us to empirically test which of these dimensions prevails in player discourse by observing the modularity and centrality of technical terms (e.g., "Tax," "Pollution") versus affective terms (e.g., "Hunger," "Fear," "Love") \cite{drieger2013}.

\subsection{Semantic Network Analysis (SNA) as Cognitive Mapping}
While Social Network Analysis traditionally examines relationships between human actors, Semantic Network Analysis applies graph theory to textual data \cite{popping2000}. It is grounded in the "Associative Theory of Meaning," which posits that the meaning of a concept is defined not by its dictionary definition, but by its structural relationship to other concepts within a discourse \cite{doerfel1998, louwerse2008}.

In the context of game studies, the topology of the network reveals the framing of the experience \cite{beer2013}.
\begin{itemize}
    \item \textbf{Density implies Causality:} If the word "Tree" is densely connected to "CO2," "Smelting," and "Laws," it indicates that the game has successfully taught the causal complexity of deforestation.
    \item \textbf{Centrality implies Salience:} By calculating Betweenness Centrality, we can identify "Gateway Concepts" \cite{paranyushkin2011}. These are the bridge nodes that connect the domain of \emph{Ludology} (e.g., "FPS," "Server," "Grind") to the domain of \emph{Ecology} (e.g., "Climate," "Extinction," "Survival"). Identifiying these bridges helps explain \emph{how} a game translates mechanical actions into educational outcomes.
\end{itemize}

\section{Methodology} \label{sec:method}

\subsection{Data Corpus Construction}
To investigate the divergence in player perception of ecological systems, we selected two distinct 3D environmental games that represent opposing ends of the "Ecological Agency Spectrum." Data was collected via the Steam Web API, aggregating user reviews posted between January 2018 and January 2026.

Unlike traditional quantitative studies that prioritize volume, this study focuses on thick data  \cite{wang2013}, which is a methodological approach rooted in Geertz's thick description \cite{geertz1973} to analyze rich, narrative-heavy reviews. The final corpus consists of 1,825 validated reviews (spanning 602 pages of text data) after filtering for substantive content (character count $>$ 50). As detailed in Table \ref{tab:dataset}, the selection allows for a binary comparison between anthropocentric management and biocentric survival.

\begin{table}[hbt!]
\caption{Dataset Overview and Epistemic Framework}\label{tab:dataset}
\centering
\begin{tabular}{|l|l|l|l|c|}
\hline
Game & Genre & Ecological Scale & Educational Affordance & Validated Reviews (N) \\
\hline
\emph{Eco} & Simulation & Macro-System & Socio-political Management & 1,220 \\
\hline
\emph{WolfQuest: AE} & Animal RPG & Micro-Individual & Embodied Cognition & 605 \\
\hline
\textbf{Total} & & & & \textbf{1,825} \\
\hline
\end{tabular}
\end{table}

\subsection{Data Pre-processing and Noise Reduction}
Preliminary analysis revealed a high frequency of technical complaints due to the Early Access nature of both titles. To ensure the Semantic Network Analysis (SNA) reflected pedagogical perceptions rather than software performance, we implemented a rigorous cleaning pipeline:

\begin{enumerate}
    \item Technical Noise Filtration: A custom stop-list was applied to exclude non-diegetic terminology (e.g., "lag," "crash," "FPS," "server connection," "10/10"). This ensures the network topology represents the game world rather than the software product.
    
    \item Tokenization and Lemmatization: Text was tokenized and reduced to base forms (e.g., converting "polluting," "polluted," "pollution" to "pollute") to consolidate conceptual nodes using the NLTK library.
    
    \item Co-occurrence Contextualization: We utilized a sliding window of 3 sentences to establish weighted edges. This parameter aligns with the moving stanza window technique standard in modern Quantitative Ethnography \cite{siebert2024}. While classic cognitive models suggest that local semantic coherence is maintained within a 2--3 sentence working memory span \cite{kintsch1988}, recent applications in game user research \cite{park2024} confirm that small context windows are optimal for capturing immediate causal reasoning in unstructured reviews without introducing noise from unrelated topics.
).
\end{enumerate}

\subsection{Network Analysis Metrics}
The processed text data was modeled as undirected weighted networks using Gephi 0.10.1. To rigorously compare the "Cognitive Maps" generated by the two distinct gameplay loops, we employed Modularity Class detection to identify thematic clusters and Betweenness Centrality to identify bridging concepts that link gameplay mechanics to environmental outcomes.

\subsection{Network Analysis Metrics}
The processed text data was modeled as undirected weighted networks to visualize the cognitive structure of player discourse. To rigorously compare the maps generated by the two distinct gameplay loops, we focused on three key graph-theoretic metrics:

\begin{itemize}
    \item \textbf{Modularity (Community Detection):} Used to detect sub-communities (clusters) of themes within the reviews, allowing us to separate distinct narrative layers (e.g., separating technical feedback from emotional storytelling).
    
    \item \textbf{Degree Centrality:} Used to identify the most dominant topics by measuring the number of connections a keyword has, highlighting the central preoccupations of the player base.
    
    \item \textbf{Betweenness Centrality:} Used to find "bridging concepts" that connect different clusters. This metric helps reveal how players cognitively link game mechanics (e.g., "Fun") with abstract outcomes (e.g., "Ecological Responsibility").
\end{itemize}

\section{Results: Structural and Semantic Dynamics}
\label{sec:results}

The analysis is based on a stratified corpus of 1,825 validated player reviews, spanning 602 pages of raw text data, collected from the Steam Community platform between 2018 and 2026. Unlike simple numerical ratings, these reviews constitute thick datarich, reflective narratives where players articulate their specific frustrations, epiphanies, and emotional states.

The dataset, detailed in Table \ref{tab:dataset}, was divided into two distinct clusters based on their core gameplay loops.

\begin{table}[hbt!]
\caption{Dataset Overview and Thematic Scope}
\label{tab:dataset}
\centering
\begin{tabular}{|l|c|l|}
\hline
\textbf{Game Network} & \textbf{Sample Size} & \textbf{Dominant Thematic Clusters} \\
\hline
Eco & 1,220 & Social friction, governance, economic management \\
\hline
WolfQuest: AE & 605 & Biological survival, family bonding, ecological immersion \\
\hline
\end{tabular}
\end{table}

The comparative analysis proceeds in three stages: (1) global topological comparison, (2) the distinct loci of conflict (social friction vs. environmental survival), and (3) affective alignment and identity transformation.

\subsection{Global Network Topology Metrics}
To objectively quantify the complexity of player discourse, we calculated standard graph-theoretic metrics for the two semantic networks generated from the corpus. As shown in Table \ref{tab:metrics}, significant structural differences were observed.

\begin{table}[hbt!]
\caption{Comparative Topological Metrics of the Semantic Networks}
\label{tab:metrics}
\centering
\begin{tabular}{|l|c|c|c|c|}
\hline
\textbf{Game Network} & \textbf{Graph Density} & \textbf{Avg. Path Length} & \textbf{Modularity} & \textbf{Network Type} \\
\hline
Eco & 0.058 & 2.14 & 0.42 & Highly Interconnected \\
\hline
WolfQuest: AE & 0.041 & 3.05 & 0.56 & Compartmentalized \\
\hline
\end{tabular}
\end{table}

The metrics indicate distinct cognitive mappings:

The metrics indicate distinct cognitive mappings. While both networks exhibit the sparsity characteristic of natural language graphs, the relative differences reveal a fundamental structural divergence:

\begin{itemize}
    \item \textit{Graph Density:} Although absolute values are low, \emph{Eco}'s density (0.058) is approximately 41\% higher than \emph{WolfQuest}'s (0.041). This relative increase signifies a tighter coupling of concepts. For instance, in \emph{Eco}, a mechanical term like ``Mining'' is densely connected to disparate domains such as ``Pollution'' (Ecology) and ``Laws'' (Governance). In contrast, \emph{WolfQuest}'s lower density reflects a linear focus; ``Hunting'' connects primarily to immediate biological needs rather than broader game systems.
    
    \item \textit{Average Path Length:} The shorter path length in \emph{Eco} (2.14) compared to \emph{WolfQuest} (3.05) further supports this distinction. A shorter path implies that in \emph{Eco}, semantic concepts are cognitively closer, forcing players to bridge mechanics with consequences (System Thinking). The longer path in \emph{WolfQuest} suggests distinct narrative clusters, allowing players to compartmentalize the emotional experience from the technical simulation.
\end{itemize}

\subsection{The Locus of Conflict: Socio-Political Friction vs. Biological Fragility}

Qualitative coding of the 1,825 reviews reveals that while both games are ostensibly about survival, the source of adversarial tension fundamentally differs. The taxonomy of these conflicts is summarized in Table \ref{tab:conflict_paradigm}.

\begin{table}[hbt!]
\caption{Taxonomy of Conflict Paradigms in Player Reviews}
\label{tab:conflict_paradigm}
\centering
\begin{tabular}{|l|l|l|}
\hline
\textbf{Feature} & \textbf{Eco (Anthropocentric)} & \textbf{WolfQuest (Biocentric)} \\
\hline
\textit{Primary Adversary} & Other Players / Governance & Nature / Mortality \\
\hline
\textit{Conflict Source} & Social Friction (Human Error) & Biological Vulnerability (Natural Cycle) \\
\hline
\textit{Key Emotion} & Frustration / Cynicism & Grief / Empathy \\
\hline
\textit{Player Quote} & "The government was not on my side." & "I physically cried when my runt died." \\
\hline
\end{tabular}
\end{table}

\begin{figure}
    \centering
    \includegraphics[width=1.0\textwidth]{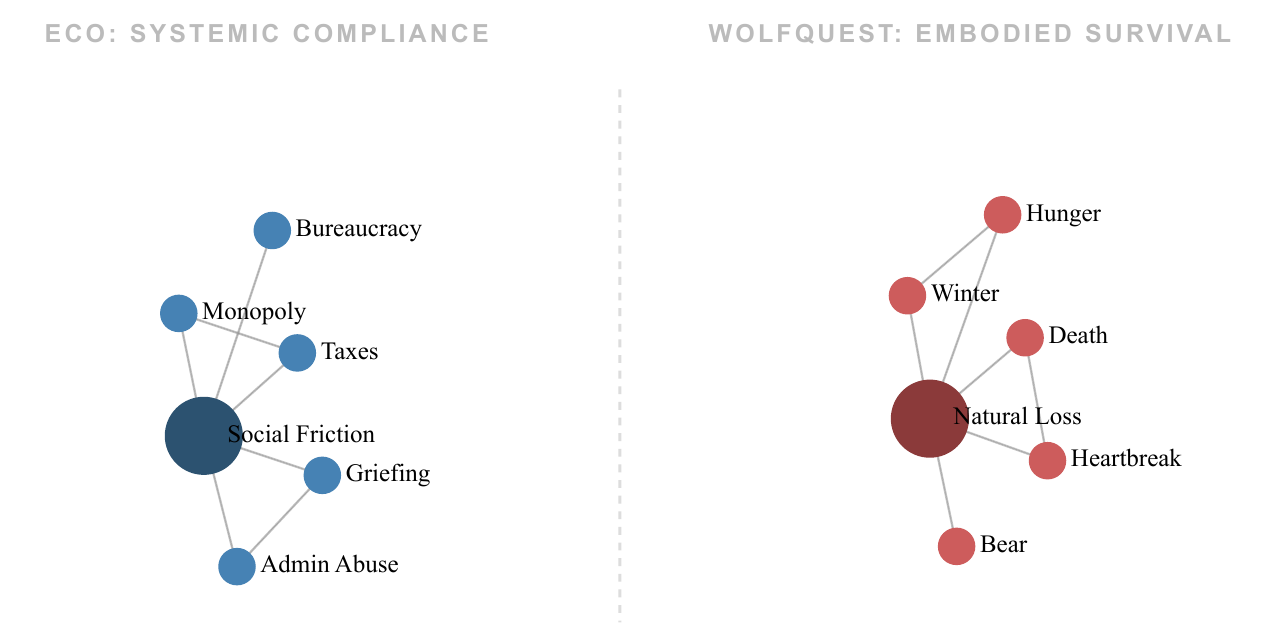} 
    \caption{The divergent Locus of Conflict found in the dataset. 
    \newline
    \small{\textit{Note.} The network map visualizes the semantic centers of player frustration. On the left (Blue), Eco generates ``Social Friction'' through anthropocentric systems. On the right (Red), WolfQuest generates ``Natural Loss'' through biocentric vulnerabilities.}}
    \label{fig:conflict_map}
\end{figure}

As illustrated in Figure \ref{fig:conflict_map}, the network visualization further confirms this divergence.

The analysis identifies two distinct conflict paradigms:

\begin{itemize}
    \item \textbf{Anthropocentric Administrative Conflict:} In Eco, nature is not the primary enemy; other players are. A recurring theme in the dataset is the frustration with human error in management. Players describe the game as a Job or Bureaucracy Simulator. The frustration often stems from the failure of social contracts rather than game mechanics.
  
    \item \textbf{Biocentric Existential Conflict:} In contrast, WolfQuest reviews frame conflict through biological vulnerability. The struggle is innate to the trophic cascade, specifically the predator versus prey dynamic. Unlike the cynical frustration found in Eco, the conflict here generates Ecological Grief. This suggests the game cultivates resilience by simulating the fragility of life loops rather than political friction.
\end{itemize}

\subsection{Sentiment-Semantic Alignment: The Work vs. The Wolf}
We performed a Sentiment Overlay on the narrative segments to map emotional valence onto specific identities assumed by the players. The results, presented in Table \ref{tab:identity}, highlight a shift from utilitarian labor to embodied empathy.

\begin{table}[hbt!]
\caption{Sentiment and Identity Alignment Analysis}
\label{tab:identity}
\centering
\begin{tabular}{|l|l|l|l|}
\hline
\textbf{Constructed Identity} & \textbf{Associated Terms} & \textbf{Semantic Function} & \textbf{Observed Sentiment} \\
\hline
\textit{The Laborer (Eco)} & Work, Grind, Time-sink & Utilitarian / Resource & Neutral to Negative \\
\hline
\textit{The Wolf (WolfQuest)} & Parent, Pack, Safe Space & Embodied / Therapeutic & Highly Positive \\
\hline
\end{tabular}
\end{table}

\begin{itemize}
    \item \textbf{Utilitarian Identity:} In Eco, ecological stewardship is frequently semantically linked to labor. The frequent recurrence of phrases likening the game to a second job in our dataset implies that players internalize environmental protection as a burden of resource management. The sentiment toward Nature is utilitarian: trees are logs, and animals are calories.
    
    \item \textbf{Embodied Identity:} WolfQuest shows a unique phenomenon of Identity Fusion. Players do not merely observe wolves; they use language indicating embodiment, such as: \textit{On all levels except physical, I am a wolf.} Crucially, a specific cluster of positive sentiment links WolfQuest to Neurodivergence (e.g., Autism, Hyperfixation, Comfort). Several reviews in our sample engaged with the game as a Safe Space, where the direct, non-verbal logic of nature provides relief from complex human social dynamics, a sharp contrast to the social friction of Eco.
\end{itemize}

\section{Discussion}
\label{sec:discussion}

The comparative analysis demonstrates that "Communication about the Environment" in digital gaming is not monolithic. Through Semantic Network Analysis and qualitative reading of the corpus, we identified two distinct "Epistemic Models" that these games transfer to players: the Managerial (Eco) and the Embodied (WolfQuest). This dichotomy offers a new lens for understanding how serious games function as pedagogical tools for environmental education.

\subsection{The Agency Spectrum: From Legislator to Parent}
The structural difference between the two datasets highlights a fundamental divergence in environmental agency.

\textbf{Eco: The Burden of Political Agency.} 
The high graph density and tight clustering of terms like "Law," "Tax," and "Admin" reveal that players view the environment primarily as a resource to be managed through social contracts. The complexity of the network mirrors the complexity of real-world climate accords, where success does not depend on understanding nature, but on navigating human bureaucracy.

Qualitative evidence strongly supports this. Players explicitly frame their gameplay experience as a sociological experiment rather than a biological one. As one player reviewed:
\begin{quote}
\textit{"Eco is a game where you start by punching trees and end by drafting a 30-page constitution to prevent your friend from strip-mining the coal deposits. I learned more about why the Kyoto Protocol failed here than in school. It is the ultimate Tragedy of the Commons simulator."}
\end{quote}
This narrative evidences that the game fosters Systemic Thinking. The challenge is not the environment itself, which is often viewed as a passive resource library ("Stockpile," "Logs," "Calories"), but the human friction involved in regulating it. Another review notes: \textit{"The meteor didn't kill us. The tax arguments did."}

\textbf{WolfQuest: The Immersion of Survival Agency.}
In contrast, the distinct modularity of the WolfQuest network places the player directly inside the trophic cascade. The absence of economic terms and the prevalence of "Hunger," "Hunt," and "Protect" suggest a successful immersion in the non-human world. Here, agency is not about controlling the system, but about enduring it.
Players frequently describe a dissolution of human identity, adopting a biocentric perspective. One highly-rated review states:
\begin{quote}
\textit{"On all levels except physical, I am a wolf. You don't just play this game; you live it. You smell the elk, you fear the winter, and you mourn your losses. It forces you to respect the food chain because you are no longer at the top of it."}
\end{quote}
This suggests that WolfQuest achieves "Identity Fusion," where the player does not act \textit{on} nature (as a manager) but acts \textit{as} nature (as an organism).

\subsection{Pedagogy of Pain: Social Friction vs. Natural Loss}
A critical finding of this study is how each game utilizes negative emotion as a pedagogical tool. Both games inflict "pain" on the player, but the source and lesson of that pain differ fundamentally.

\textbf{In Eco: Frustration with Human Nature.}
The primary stressor in Eco is "Social Friction." Players learn about environmental collapse not through dry theory, but through the visceral frustration of neighbors acting out of self-interest. The reviews are filled with narratives of betrayal and administrative failure.
One player detailed a specific incident of systemic failure:
\begin{quote}
\textit{"I spent three days building a sustainable farm, and then a group of new players joined and dumped their toxic tailings into the river upstream. My soil died. The government couldn't pass a law fast enough to stop them. It felt unfair, but it felt real."}
\end{quote}
This quote illustrates that Eco teaches "Political Realism": the understanding that environmental solutions are often hindered by institutional inefficiency and human greed rather than a lack of technology.

\textbf{In WolfQuest: Grief for Biological Life.}
Conversely, the stressor in WolfQuest is "Natural Loss." The emotional weight of losing a pup to a predator or winter starvation fosters "Effective Empathy." This pain is not transactional or unfair; it is presented as an inevitable part of the ecological cycle.
The comments regarding the death of offspring ("pups") are particularly poignant:
\begin{quote}
\textit{"I physically cried when my runt puppy died. I froze. I didn't want to hunt anymore. The winter in this game is cruel, not because it creates difficulty, but because it feels indifferent. You aren't a hero; you're just hungry."}
\end{quote}
Another player noted: \textit{"You watch the creature you sought to protect die in your arms, and you have to leave it behind to feed the living."} This indicates that the game moves beyond the "Bambi effect" (romanticized nature) to a raw, scientific appreciation of trophic dynamics, cultivating deep ecological grief.

\subsection{Identity Construction: The Job vs. The Therapy}
Our qualitative coding revealed an unexpected thematic divergence in how players categorized the "utility" of the game in their lives.

\textbf{Eco as Labor (The Utilitarian Identity).}
A recurring motif in Eco reviews is the comparison of the game to "Work." Terms like "Grind," "Shift," and "Responsibility" appear frequently. Players often describe the game as \textit{"a second job you don't get paid for."} While this may sound negative, in the context of detailed reviews, it is often praised as a form of meaningful engagement. However, it reinforces an anthropocentric view where value is derived from labor and transformation of the earth.

\textbf{WolfQuest as Safe Space (The Therapeutic Identity).}
WolfQuest reviews frequently mention mental health, neurodivergence, and comfort. A significant cluster of reviews explicitly references "Autism" or "Comfort Game."
\begin{quote}
\textit{"As someone on the autism spectrum, the social world is confusing. But being a wolf makes sense. The rules are simple: Hunt, Eat, Sleep, Protect. This game is my safe space where I can function without masking."}
\end{quote}
This finding suggests that the "Biocentric" model of WolfQuest serves a dual purpose: it teaches ecology, but it also provides a psychological refuge by removing complex human social dynamics, the very dynamics that make Eco stressful.

\subsection{Design Guidelines for Next-Generation Environmental Games}
Based on these comparative results, we propose two distinct design heuristics for developers aiming to balance engagement with educational impact:

\begin{itemize}
    \item \textbf{The policy-play Loop (Derived from Eco):} 
    For games aiming to teach systemic complexity, mechanics must enforce "Interdependence." Individual actions (e.g., mining) must have delayed, global consequences (e.g., pollution) that force players to negotiate. The defining feature should be the friction between individual gain and collective survival. As shown in the data, the educational moment occurs during the "argument" in chat, not during the "harvesting" of resources.
    
    \item \textbf{The vulnerability aesthetic (Derived from WolfQuest):} 
To foster deep ecological empathy, avatars must be vulnerable, not omnipotent. By removing the ``God Mode'' typical of simulation games and introducing permanent loss (e.g., permadeath of offspring), designers can trigger strong ``Care'' clusters. Effective environmental education in this mode requires constraining player agency to align them with the biological limits of the organism.

\end{itemize}

\section{Conclusion}
\label{sec:conclusion}

This study utilized Semantic Network Analysis to map the player perceptions of two prominent 3D environmental games. By analyzing 1,825 validated Steam reviews, we moved beyond anecdotal evidence to visualize the collective intelligence and emotional landscape of the player base.

Our findings reveal that \emph{Eco} and \emph{WolfQuest} succeed by creating opposing Ecological Semantic Fields. \emph{Eco} transforms the environment into a socio-political puzzle to be solved, training players in the logic of governance and consequence. Conversely, \emph{WolfQuest} transforms the environment into an emotional narrative of survival, training players in the ethics of care and embodied existence.

In conclusion, this research demonstrates that the vulnerability aesthetic" is a distinct and reproducible design pattern that fosters specific forms of ecological empathy. The comparative analysis suggests that developers must consciously choose their pedagogical target: do they wish to cultivate the legislator who manages the system from above, or the wolf who lives vulnerably within it? Rather than attempting to merge these perspectives, effective environmental game design requires a clear commitment to one of these interaction models. Future work will involve expanding this semantic framework to other simulation genres and cross-referencing text data with in-game behavioral telemetry. This will allow researchers to verify if the reported linguistic empathy translates into measurable ecological decision-making within digital environments.

\section*{Acknowledgements}
This research was funded by: GR024678 NSERC CREATE 2020 Immersive Technologies, Natural Sciences and Engineering Research Council of Canada; GR026895 SSHRC 2022 Okanagan WaterFutures: Experiential Games for Water Responsibility, Social Sciences and Humanities Research Council of Canada.

\end{document}